# The Efficiency Examination of Teaching of Different Normalization Methods

Márta Czenky

Department of Informatics, Szent István University, Gödöllő, Hungary
marta.czenky@t-online.hu

*ABSTRACT*

*Normalization is an important database design method, in the course of the teaching of data modeling the understanding and applying of this method cause problems for students the most. For improving the efficiency of learning normalization we looked for alternative normalization methods and introduced them into education. We made a survey among engineer students how efficient could they execute the normalization with different methods. We executed statistical and data mining examinations to decide whether any of the methods resulted significantly better solutions.*

## 1. INTRODUCTION

Normalization is an important method of the design of relational databases. In the course of the normalization we determine the functional and multivalued dependences in the table then we ensure the match to normal forms with decomposition of the tables. The method is a theory based strict planning method with many advantages:

- the planning process is adaptable,
- it eliminates redundancy,
- it eliminates to insertion, modification and deletion anomalies,
- it results saving of more space in storing,
- you may add new tables to the database and new rows to the table without any difficulty,
- it ensures data consistency,
- it ensures referential integrity,
- after normalization you may execute data control in the database more easily, [1], [2], [3], [4], [5].

From the disadvantages of the normalization you have to mention that it results more tables with low number of columns, moreover after normalization the query from more table can lead to complicated join operation. We can say that it is a disadvantage that the learning of the algorithm is difficult, [2], [3].

You can cascade the normalization procedure into steps. You should have such tools for the teaching of the topic which enables students to practice these steps on demand.

Several interactive computational educational systems were developed for teaching normalization, a part of them have the normalization practiced trough test solution while the others execute the normalization.

- NORMIT – was developed at University of Canterbury, Christchurch, Australia, it works on web surface nowadays too. After determination of functional dependencies, closure and primary key the students have to decompose the tables until fruition of Boyce-Codd normal form. The system compares the student's solution with ideal solution generated by problem solver and sends feedback, [6], [7].

- E-learning tool teaching dependency theory – the system was developed by the fellow workers of the University of Westminster, London and the King's Collage, London. It runs under UNIX operating system and was written in Java. First you have to give the attributes of examined relation for the program, then the determinant and dependent attributes of the functional dependencies. Then the student can choose until which normal form should the program decompose the relation. After the running the appropriate algorithm the program provides the decomposed relations and signs the primary keys, [8], [9].

- Database normalization tool – it was developed at Cornell University in 2003, it runs on web surface. After the giving relation's attributes and functional dependencies the system determines the global functional dependencies and signs which normal forms are violated by them. Afterwards, by request the program decomposes the relation and examines if the decomposition is lossless and dependency observer, [10].

- E-learning tool with use of alternative normalization method – this application written in Java is based on normalisation method named cookbook. The program executes the normalization until $3^{rd}$ NF if the functional dependencies are not trivial, that is there are not common attribute on left and right hand side, and if the functional dependencies are closed that is the right side includes all attributes which are determined by leftward attributes, [11].

- Gradiance – the system is used at Stanford University, the curricula is the content of several books the students get tasks related to them. For the verification of the understanding of normalization there are different worksheets in functional dependencies, multivalued dependencies and normalization topics. After sending worksheet the system evaluates it and gives longer explanations on wrong answers, [12].

- RDBNorma – this tool was developed by researchers of Vishwakarma Institute of Technology and Pune Vidhyarthi Griha's College of E&T. This tool stores functional dependencies in a linked list and executes relational database schema normalization up to third normal form [13].

Beside the computational education systems there are many ways to support the learning of normalization. [14] and [15] introduce the AER (Articulated Entity Relationship) diagram which is an modified E-R diagram and includes functional dependencies between attributes. [16] introduces concept maps which describe relationships between concepts of relational model, functional dependencies and normalization. [17] emphasizes the importance of usage of practical examples in course of teaching of normalization. In order to improve the efficiency of teaching of normalization you can introduce new normalization methods into education, for example [18] introduces a new normalization algorithm with which you can execute normalization up to third normal form.

In the course of teaching database design the teaching of normalization cannot be avoided. It causes a lot of problems for students based on our teacher experience. Among other things the followings cause problems:

- to determine functional dependencies, candidate keys and primary keys,
- to determine which normal forms are violated by functional dependencies,
- the strict compliance of the planning steps,

- the consequent application of the table's decomposition algorithm,
- the designation of the primary and foreign key in the decomposed tables.

We looked for alternative normalization methods and introduced them into education to improve the efficiency of the learning this method. After teaching of data modeling we made a survey among mechanical engineering and environmental engineering students to see with which method are they able to normalize more effectively.

## 2. QUESTIONNAIRE SURVEY

In 2008 we made a questionnaire survey among mechanical engineering students, in which among other things we asked the students to designate what caused problems for them at the area of database management. We compiled the questionnaire by the instructions of [19]. We asked all students of the courses to fill in the questionnaire (63 students), 54 of them gave an answer. As students happened not to fill in the questionnaire randomly we can consider the survey representative.

The data of the Table 1 show that how many percentage of students found the concepts, algorithms and activities of the main topics of the database management problematic. According to the judgment of students the most difficult topic is the normalization, medium difficult is the use of SQL language and least problematic is the E-R modelling, [20].

Table 1. The proportion of students who marked problem at the topics of database management

| Activity | The proportion of the students who marked problem |
|---|---|
| Making E-R model | 19.6% |
| Transcribing E-R model into relational model | 21.7% |
| Normalization | 69.6% |
| Observing and application of the rules of SQL language | 39.1% |
| Creating SQL statements | 43.5% |

Table 2 shows the results of the survey made at the University of Ulster, [21]. The source did not tell how many students joined the survey and how was the sample chosen.

Table 2. The result of survey at the University of Ulster

| Topic | Very difficult | Difficult | Easy | Very easy |
|---|---|---|---|---|
| Introduction to database management | | 7.7% | 76.9% | 7.7% |
| E-R modeling | | 48.7% | 48.7% | 2.6% |
| Normalization | 12.8% | 71.8% | 12.8% | 2.6% |
| Relational modeling | | 71.8% | 25.6% | 2.6% |
| SQL | 2.6% | 41% | 48.7% | 5% |

The data of the table show that normalization was judged as the most difficult activity by students, in total more students than in our survey. They judged relational modeling difficult too. According to the opinion of nearly half of the students the entity-relationship modeling was difficult, but same number of students judged it easy. The SQL language was judged difficult or very difficult by slightly more than 40% of the students, the learning of it was easy for the other part of the students.

In the survey executed in 2008 we also asked the students which steps of the normalization caused them the biggest problems. Table 3 shows their aggregated answers, [20].

Table 3. The students' judgment of activities related the normalization

| Activity, rule | The proportion of the students who marked problem |
|---|---|
| the writing of the non first NF table | 41.3% |
| the recognition of the functional dependencies | 28.3% |
| the recognition of the multivalued dependencies | 19.6% |
| determination of the primary key | 39.1% |
| to determine the correspondence to normal forms | 69.6% |
| decomposition rule of the tables | 26.1% |
| drawing dependency diagram | 30.4% |

By the data of Table 3 the understanding of the normal forms and the recognition of the level of normalization of the tables are the most difficult activities for students. The recognition of the dependencies, the drawing of a dependency diagram and the decomposition of the tables by decomposition rule are the least of all problematic activities.

## 3. METHODS OF NORMALIZATION

We teach regularly two methods of the normalization from the methods introduced hereinafter, these are the conventional normalization method and the normalization based on dependency diagram. The students also solve regularly the normalization test made by us in the course of individual preparation. We do not teach the $4^{th}$ normalization method, the cookbook method to students during the class, because we consider it automatic which can be executed without thinking. We still introduce it yet so that students can use it in the course of their latter work if they have to execute correct normalization.

### 3.1 The Conventional Method of Normalization

We don't describe the conventional method of the normalization via an example we only summarize the related knowledge because it can be found in every text book dealing with database management and well known for every professionals dealing with informatics.

The functional dependency is the unambiguous definition one of the table's columns by other columns of the table. In case of multivalued dependency the values of a table column define a set of values in another table column and these values are independent from the values of the other columns of the relation.

We teach the normal forms unto $4^{th}$ NF, the specifications of the normal forms are the followings:

$1^{st}$ NF     atomic value, primary key, no recurring group,

$2^{nd}$ NF     no partial dependency,

$3^{rd}$ NF     no transitive dependency,

BCNF     all determinant attribute is super key,

$4^{th}$ NF     no multivalued dependency, [4], [5].

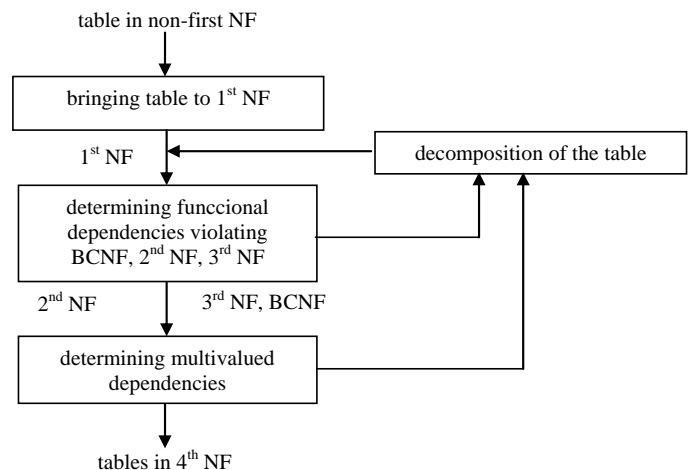

Figure 1. The process of the normalization

In the course of the normalization we determine the functional and multivalued dependencies existing in the table then we ensure the correspondence to normal forms with the decomposition of the table, see Figure 1.

## 3.2 Normalization Based on Dependency Diagram

In favor of better overview you can depict the functional dependencies existing in the several tables on a diagram. On a diagram the layout of the table's column may be horizontal or vertical. [22] gives an example for the first layout, while you can see the second layout in the book of [23]. The disadvantage of the horizontal layout is that due to lack of space you may have to visualize the dependencies in more rows.

We apply the vertical layout and we visualize both functional and multivalued dependencies on diagram. We draw the dependency diagram according to followings:

- We indicate the functional dependency with arrow, the multivalued dependency with double arrow. We include the attributes in boxes an arrow goes from the determinant attribute to the dependent attribute.

- Every attribute appears only once in the diagram.

- We do not mark the transitive dependency neither the dependency from the alternative keys, [24].

For the completion of the normalization in the left hand side column of the diagram we depict columns of such a super key which include all the determinant attributes of all dependencies. On diagram we depict the dependencies existing between the attributes of the super key too.

The process of the normalization does not change even if we do it based on dependency diagram. We take out the dependencies violating normal forms into an independent table and the determinant attribute will be the primary key in the new tables. The dependent attributes have to be deleted from the diagram, what we mark with crossing the attributes, so the diagram shows always how the original relation changes. The determinant attributes remain on the diagram and in the original table as foreign keys.

The takings out can be classify into two groups, first we take out the dependencies in which the dependent attributes do not belong to the super key then we take out the dependencies existing between the attributes of the super key.

In the first group the order of the taking out is not necessary to be considered it can be for instance $3^{rd}$ NF – $2^{nd}$ NF – BCNF. It does not have to be started with taking out the dependency violated BCNF because if the determinant attributes of this dependency depend from further attributes these latter dependencies appear between the dependencies of the attributes of the super key and they remain after taking out the dependency violated BCNF on diagram too. It can be easily proved that $2^{nd}$ NF –$3^{rd}$ NF or $3^{rd}$ NF – $2^{nd}$ NF taking out orders result the same tables. We apply the usual taking out order at the dependencies existing between attributes of super key, [24].

### 3.3 Normalization with Solving Test

In the education we have been using Moodle e-learning education system since 2007, which allows the development of question bank and solving computational tests. This system gives excellent chance to practice normalization in a way that certain questions go through the steps of the normalization as the following:

1. Recognition of the functional and the multivalued dependencies.
2. Determination of the primary key after the determination of the closure of the set of attributes.
3. To determine which normal forms are violated by functional dependencies.
4. To determine in which normal form is the table before the decomposition.
5. The decomposition of the table.
6. Determination of the primary keys of the decomposed tables.
7. Determination of the foreign keys of the decomposed tables.

Students gladly solve the computational tests it is great help for them that the order of the question takes them through the algorithm of the normalization. We elaborated more practicing series of tasks and there is normalization task in the exam tests too.

### 3.4 Normalization by Cookbook Method

For applying of this method you have to determine the functional dependencies existing in the table. Then the steps of the algorithm are the following:

1. Let us count the attributes of the functional dependencies on the left hand side (LHS) and right hand side (RHS) and let us write them beside the sides!
2. The LHS stays always untouched. We do not ever delete the attributes on the left hand sides not even if another functional dependency contains this attribute on the left hand side.
3. After this you have to delete the repetitive attributes from right hand side in a way that every attribute occur only in one dependency on right hand side.
    a. Let us keep the attributes in those dependencies where LHS belonging to them is lower and let us delete another occurrence of the attributes!
    b. If the LHS values of two dependencies are the same let us keep the attributes in those dependencies to which lower RHS belongs and let us delete the other occurrences!
    c. Let us convert the dependencies into relations the primary keys of the tables will be the attributes on left hand sides, [25], [26]!

## 4. NORMALIZATION SURVEY

At Szent István University we teach database management at more departments in BSc and MSc qualifications too, among others for mechanical engineering and environmental engineering students. The two main topics of the subject are data modeling and the SQL-92 language. In the course of the teaching of data modeling we introduce database design with the help of E-R modeling and normalization too.

Table 4. The characteristics of taught subjects

| Name of subject | Abbreviation | Major | Qualification/ Course | Class per week lecture+ practise | Year | Head count |
|---|---|---|---|---|---|---|
| Computer Studies III. | KM3 | environmental engineer | BSc, full time | 2+2 | 2013 | 37 |
| Database Management | ABK | mechanical engineer | BSc, full time | 0+2 | 2010 | 21 |
| Environmental Databases | KDB | environmental engineer | MSc, full time | 2+2 | 2013 | 9 |

We had a survey done among the students of one course in 2010 and two courses in 2013 to determine which normalization method takes students to more effective normalization.

In Table 4 we summarized the characteristics of the subjects taught in courses participated in survey. The curriculum of database management is the same in the case of all tree subjects. In near equal (time) ratio we teach data modeling (E-R and relational model, normalization) and SQL-92 language, [4], [5]. At the four classes per week courses students attend database management in the first half of the semester, while in the second half of the semester the KM3 students learn CAD and the KDB students learn about environmental databases.

The students of ABK course could decide if they participate in survey. Among them, 10 students solved the normalization tasks. According to the homogeneity test executed by the distribution of marks in the end of the semester this pattern has same distribution than full course. All KDB students participated in survey. At KM3 course only 20 students executed the normalization with cookbook method. We verified with Chi-square test if this pattern does have same distribution than the course. We found that they have same distribution. All students of KM3 course executed the normalization with other three methods. You can consider the results of all patterns as representative. Students had to execute the normalization with different tasks by the following four methods:

- conventional normalization method,
- normalization based on dependency diagram,
- normalization with solving test questions,
- normalization by cookbook method.

They had to normalize tables in which existed only functional dependencies violated second and third normal forms. We proceeded so because the recognition of the functional dependencies violated BCNF and multivalued dependency were difficult for the students and our goal was that they could solve the tasks.

## 5. INVESTIGATION METHODS

First we review those activities at all courses which were executed non adequately by students that is the average results were fewer than 50% and those activities which were executed well by them that is the students achieved near 90% result.

For every course, in case of all methods we calculate the means and standard deviations then we examine with Chi-square test if the results have normal distribution.

Then we compare the result of conventional method with the result of the other three methods in pairs to decide if the students can normalize with some alternative methods significantly better than with conventional method.

As the results of different methods are not independent, we verify with T-test for dependent samples the homogeneity of the standard deviations. If they are the same, then we verify with two samples paired T-test if the means are the same. If the standard deviations are not same then we verify with Wilcoxon-test if the means are same. In this way we can decide if the results have same distribution or the students achieved significantly better result with some alternative method than with conventional method, [27], [28], [29].

We also examine if we can find some connection between results of different normalization methods. For this reason in the highest headcount KM3 pattern we determine correlation coefficients between the result of conventional method and the results of alternative methods. We make significance test to decide if the correlation coefficients differ from zero.

At KM3 course we depict the results of alternative methods in function of result of conventional method on scatter chart. We hope that this chart will show with which method the students can normalize better. We determine two-two clusters by methods which reflect the average results better than means. We also determine how many percent of students achieved better result with alternative methods than with conventional method.

We search for decision trees and associations rules in results of KM3 course, [30]. For the examination we expressed the achieved results by marks of 1 to5. The worst mark is 1 while the best one is 5. The certain marks correspond to the following results: 1 (0%-50%), 2 (50,01%-62%), 3 (62,01%-76%), 4 (76,01%-86%) and 5 (86,01%-100%). At the result of cookbook method we mark with zero if the students did not participate in the pattern. With these examinations we look for an answer for the question that in case of a given result by conventional method which results can we expect with alternative methods.

## 6. THE RESULTS OF THE EXAMINATIONS AND THEIR ASSESSMENT

We do not review the results of ABK and KDB courses in details because these patterns have lower headcount, we only summarized the results.

At both courses those activities are near the same in which with one or more methods the students achieved worse result than 50%:

- in which normal forms is the given table,
- writing of table remaining after take out,
- designation of foreign key in decomposed tables.

The last two activities were problematic the most at those the results were bad by more methods.

It is great to see that students achieved the best result in the recognition of functional dependencies and the taking out of functional dependencies violating $2^{nd}$ and $3^{rd}$ normal forms.

We checked if the results have normal distribution. We found that all four results have normal distribution at both courses. In case of all methods the means indicate medium efficiency. At both courses the mean of the result of the normalization based on dependency diagram is the highest and its standard deviation is the lowest, while the means of the result of normalization with cookbook methods is the lowest and its standard deviation is the highest.

The students met with the cookbook method first in the survey. We described the method on the worksheet and gave a sample of its use. Students had to interpret and understand it than they had to solve the appointed task accordingly. Perhaps this is the reason of the worst results of this method.

The executed statistical tests by pairs show that the distribution of result of alternative normalization methods is same than the distribution of the result of conventional normalization method. The differences in means and in standard deviations are by coincidence. Thus we cannot say that the result of normalization based on dependency diagram is significantly better than the result of conventional method.

Table 5. Results fewer than 50% KM3 course

| Activity | Method | | | |
|---|---|---|---|---|
| | Conventional | Dependency diagram | Test | Cookbook |
| determination of primary key | | | 33.3% | |
| recognition of functional dependencies violated 2$^{nd}$ NF | | | | 40% |
| recognition of functional dependencies violated 3$^{rd}$ NF | 43.2% | | | 32.5% |
| taking out of functional dependencies violated 2$^{nd}$ NF | 44.6% | | | |
| writing of table remained after taking out | 28.4% | | | |
| designation of foreign key in decomposed tables | 27% | 48.6% | | 47.5% |

In Table 5 we represented those activities of KM3 students in which the result was lower than 50%. Apparently the recognition and taking out of functional dependencies are problematic and henceforward the most difficult activity is the designation of foreign key in decomposed tables.

At conventional method the results of those activities which are not included in Table 5 are between 50%-70% while at normalization based on dependency diagram the results are between 75%-97%. This indicates that students can normalize better with second method. The results of the other two alternative methods are same than result of conventional method.

Table 6. Means and standard deviations KM3 course

| Feature | Method | | | |
|---|---|---|---|---|
| | Conventional | Dependency diagram | Test | Cookbook |
| has normal distribution? | yes | yes | yes | yes |
| mean | 4.6 | 5.4 | 5.27 | 4.67 |
| standard deviation | 1.87 | 0.7 | 1.47 | 1.55 |
| achieved better result than with conventional method | | 91.81% | 62.16% | 78.95% |

The obtainable maximal score were 7 at every task. We summarized the means and standard deviations in Table 6. In our examinations the results of all four methods have normal distribution.

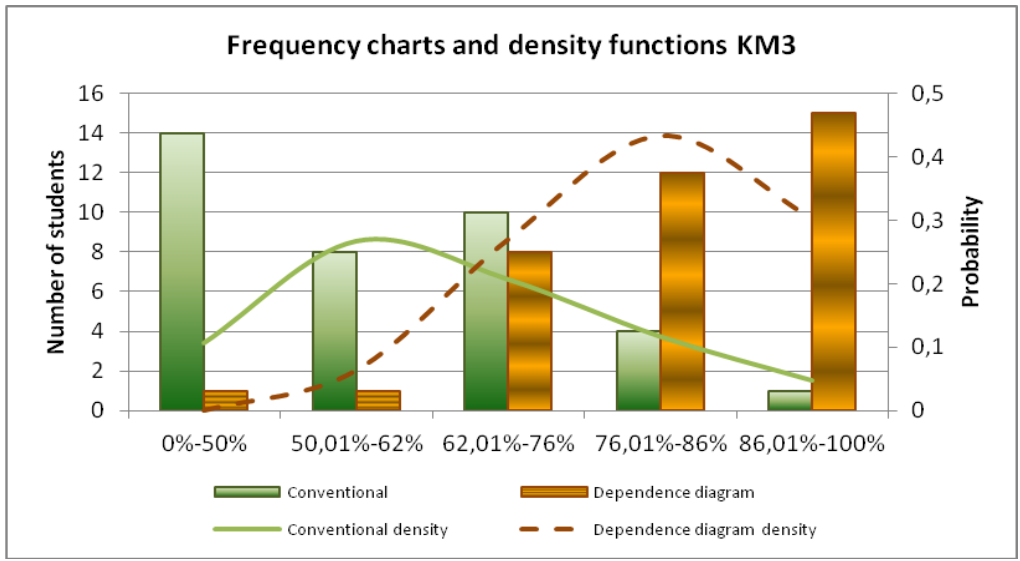

Figure 2. Frequency charts and ideal density functions KM3

You can see from Table 6 that students achieved the best result in the course of normalization based on dependency diagram while they achieved the worst result in conventional normalization method. We depicted on common chart the frequency distribution of the result of normalization based on dependency diagram and conventional method, see Figure 2.

On Figure 2 we visualized the ideal density functions calculated by the medium values of score boundaries, empirical means and standard deviations. You can read the frequency values from left side axis the values of density function from right side axis. Figure 2 also shows that the result of normalization based on dependency diagram is far better than the result of conventional normalization.

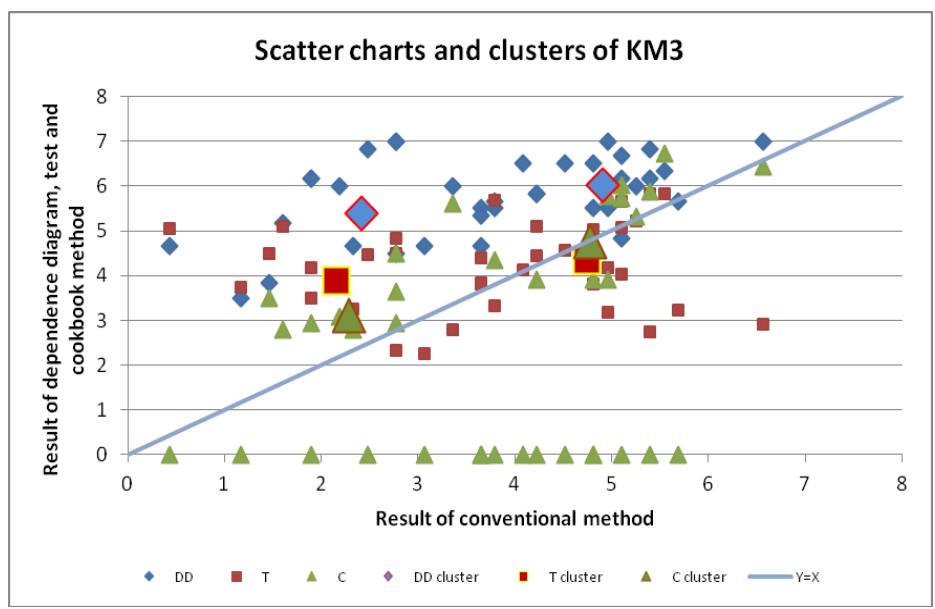

Figure 3. Scatter charts and clusters KM3

On Figure 3 we depicted the results of alternative methods as a function of the result of conventional normalization method. On the figure we depicted Y=X line too. The markers appearing above the line indicate that the students achieved better result with given alternative method than with conventional method.

We determined two-two clusters by series which show the distribution of results better than the means. At cookbook method at determination of the clusters we took into consideration the results of those students who solved the task – we disregarded zero values. On the figure the markers appearing on X axis show what result was achieved at conventional method by those students who did not solve the task of cookbook method.

Figure 3 also shows that students achieved the best result in normalization by dependency diagram, followed by test solving and cookbook method which results are near same.

Table 7. Homogeneity examination of distributions KM3

| Feature | Examination methods and results | | |
|---|---|---|---|
| | Conventional - Dependency diagram | Conventional - Test | Conventional - Cookbook |
| standard deviation | T-test for dependent samples<br>not same | T-test for dependent samples<br>not same | T-test for dependent samples<br>same |
| mean | Wilcoxon-test<br>not same | Wilcoxon-test<br>same | two-samples paired T-test<br>not same |

At KM3 course we examined too if the distribution of the results of conventional methods are the same as the distribution of the result of any alternative methods. Table 7 shows the result of the examination. According to the data of the table the distributions are not the same. As at normalization with conventional method and normalization based on dependency diagram neither the standard deviation nor the mean are equal and the mean of the second method is higher and its standard deviation is lower we can say that the result of normalization based on dependency diagram is significantly better on 0.05 significance level than the result of normalization with conventional method.

We calculated the correlation coefficients between result series of KM3 course in order to determine if there is any connection between series and how strong it is. Table 8 includes the coefficients. We verified with significance test if the correlation coefficient differs from zero.

Table 8. Correlation coefficients KM3

| Feature | Conventional - Dependency diagram | Conventional - Test | Conventional - Cookbook |
|---|---|---|---|
| correlation coefficient | 0.4464 | 0.0681 | 0.8247 |

We found that the correlation coefficient between results of conventional and test method in fact is zero therefore there is no connection between these two series. There is medium strong connection between results of normalization with conventional method and based on dependency diagram. The third data of Table 8 indicates stronger connection. The better a student solved the conventional normalization task the better he executed the normalization with cookbook method.

With the following examinations we looked for an answer for the question that in case of a given result achieved with conventional method which result can be expected with alternative methods.

We searched decision trees between results of conventional and alternative methods see Table 9 and Table 10. The tables include those results at which the confidence or the support is higher than 40%. The program cannot make decision tree between results of conventional and cookbook method.

Table 9. Decision tree between results of normalization with conventional method and based on dependency diagram KM3

| Achieved and expected result | | Confidence | Support |
|---|---|---|---|
| Conventional | Dependency diagram | | |
| 1 | 5 | 35.7% | 37,8% |
| 2 | 4 | 75% | 21.6% |
| >2 | 5 | 60% | 40.5% |

Table 10. Decision tree between results of normalization with conventional method and test method KM3

| Achieved and expected result | | Confidence | Support |
|---|---|---|---|
| Conventional | Test | | |
| <=3 | 3 | 40.6% | 86.5% |
| >3 | 1 | 60% | 16.5% |

According to data of Table 9 in case of any result of conventional method you can expect good result in normalization based on dependency diagram. The data of Table 10 show that who executed normalization with conventional method better those solved the test worse.

Table 11. Associations rules KM3

| | Confidence | Support |
|---|---|---|
| **Conventional=>Dependency Diagram** | | |
| 2=>4 | 66.67% | 16.21% |
| **Conventional=>Test** | | |
| 5=>1 | 100% | 2.7% |
| **Conventional=>Cookbook** | | |
| 2=>0 | 54.5% | 16.2% |
| 5=>4 | 50% | 2.7% |

We also looked for associations rules, see Table11. At the found rules the support values are very low which is caused by the low number of students in the pattern. In the table the first and the second rules are in line with the data of the decision trees. The next to last rule indicates that those students did not solve the normalization task with cookbook method who achieved mark 2 at conventional normalization. Indeed, they were the majority, but the markers on X axis of Figure 3 indicate that even more students did not solve this task. The last rule shows that who achieved the best result with conventional method those achieved somewhat worse result with cookbook method.

## 7. SUMMARY

By our teacher experience we may say that the introduction of discussed alternative normalization methods into education proved to be a good decision. At one course – KM3 - we could prove that the normalization based on dependency diagram provided significantly better results than the conventional method. Further advantage of introducing alternative normalization methods into education is that students could choose the most appropriate method and they could use it for solving their tasks. We are glad that in the solving of the homework more students applied the normalization based on dependency diagram with which they solved their task immaculately.

The development of the normalization tests proved also to be a good decision, students solved the tests efficiently. According to students' opinion this way of practicing normalization is useful, they solve these tests gladly. This normalization decomposed into questions can be found in the exercise book of data modelling tasks too, [31].

In two cases – normalization based on dependency diagram and cookbook method - of the discussed methods we also have to mention the advantage that the dependency violated BCNF should not be taken out as first step in all cases and the execution of the normalization does not affect the loss of some dependency.

## APPENDIX

**Normalization Based on Dependency Diagram - Example**

Let us see the following table:

T (A, B, C, D, E, F, G, H, I, J, K, L)

The following functional and multivalued dependencies exist in the table:

A → B, C

D → E, F

C, D → G

I → J, K, L

K → L

D →→ A, H

D →→ I

To determine primary key we define the closure of {A, D, H, I} set of attributes.

{A, D, H, I}$^+_1$={A, D, H, I} due to reflexivity

{A, D, H, I}$^+_2$={A, D, H, I, B, C} due to A → B, C dependency

{A, D, H, I}$^+_3$={A, D, H, I, B, C, E, F} due to D → E, F dependency

{A, D, H, I}$^+_4$={A, D, H, I, B, C, E, F, G} due to C, D → G dependency

{A, D, H, I}$^+_5$={A, D, H, I, B, C, E, F, G, J, K, L } due to I → J, K, L dependency

While the closure include all columns of the table and it is not true for any of its subset the primary key of the table will be the A, D, H, I column combination.

The listed dependencies violate the following normal forms:

| | |
|---|---|
| A → B, C | 2$^{nd}$ NF |
| D → E, F | 2$^{nd}$ NF |
| C, D → G | BCNF |
| I → J, K, L | 2$^{nd}$ NF |
| K → L | 3$^{rd}$ NF |
| D →→ A, H | 4$^{th}$ NF |
| D →→ I | 4$^{th}$ NF |

We depicted the dependencies on a diagram, see Figure 4. On the diagram we used different line type depending on which normal forms are violated by dependency.

At the taking out of the dependency we proceed according to columns of Figure 4 from right to left, so we get to the following tables. In the tables we printed the primary keys in bold and the foreign keys in italics.

T1 (**K**, L)

T2 (**A**, B)

T3 (**D**, E, F)

T4 (**I**, J, *K*)

T5 (*D*, **C**, G)

T6 (**A**, C)

T7 (*D*, *A*, **H**)

T8 (*D*, *I*)

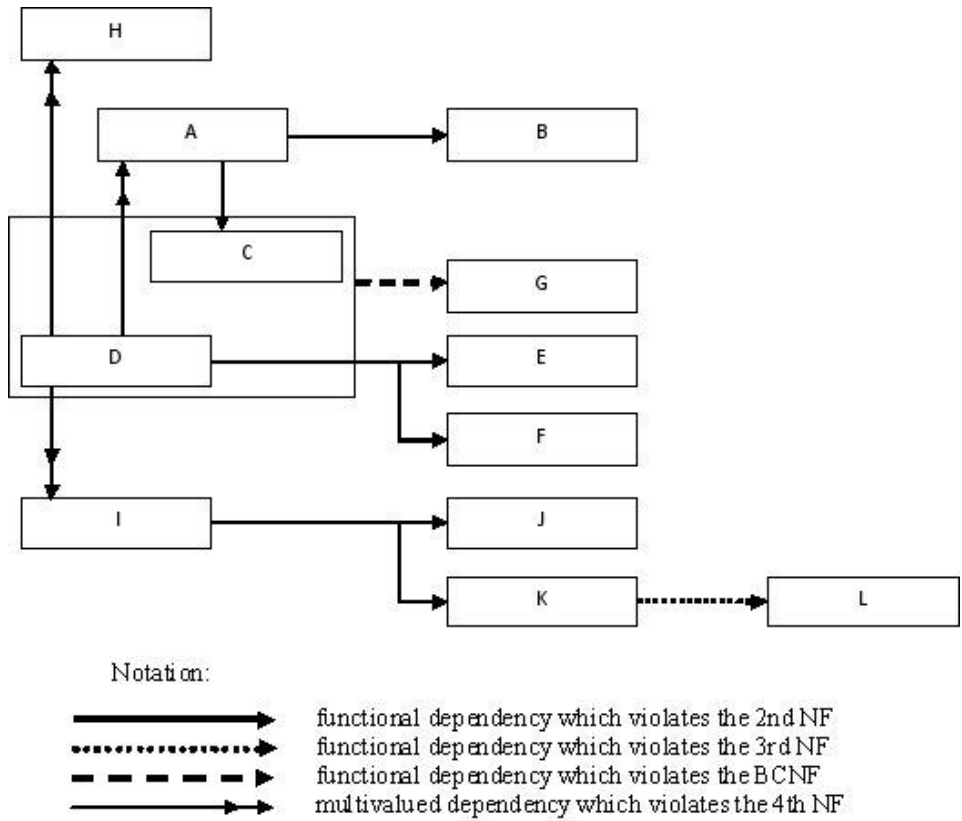

Figure 4. Dependency diagram

The 2nd and 6th tables can be merged because there have the same primary key thus we receive the following table:

T2 (**A**, B, C)

**Normalization with Solving Test - Example**

To visualize the above mentioned we introduce a question series of normalization. At question 3 and 6 the Moodle system represents the answer possibilities standing on the right hand side in a combo box and students have to choose the correct answer from it.

1. Which functional dependencies do exist in the following table?

   Rent_a_car (RegisteredNumber, CarType, ManufacturerID, ManufacturerName, RenterID, RenterName, RenterAddress, Date, Time)

   - one car is rented by one person only
   - one person may rent more cars
   - one person may rent the same car more times in another time
   - the time is given in days
   - the shortest time is one day
   - every car is manufactured by one manufacturer
   - every renter has only one address

   a) RenterID → RenterName, RenterAddress

b) RegisteredNumber → CarType, ManufacturerID, ManufacturerName

   c) RegisteredNumber, RenterID → Date, Time

   d) ManufacturerID → ManufacturerName

   e) RegisteredNumber, Date → Time, RenterID, RenterName, RenterAddress

2. What is the primary key in the following table beside the functional dependencies selected in the previous question?

   Rent_a_car (RegisteredNumber, CarType, ManufacturerID, ManufacturerName, RenterID, RenterName, RenterAddress, Date, Time)

   a) RegisteredNumber, RenterID

   b) RenterID

   c) ManufacturerID

   d) RegisteredNumber, Date

   e) RegisteredNumber

3. Which normal form is violated by functional dependencies existing in the following table?

   Rent_a_car (RegisteredNumber, CarType, ManufacturerID, ManufacturerName, RenterID, RenterName, RenterAddress, Date, Time)

   a) RenterID → RenterName, RenterAddress                               $2^{nd}$ NF

   b) ManufacturerID → ManufacturerName                                  $3^{rd}$ NF

   c) RegisteredNumber → CarType, ManufacturerID, ManufacturerName       BCNF

4. In which normal form is the following table beside the primary key determined in the second question and the functional dependencies selected in the first question?

   Rent_a_car (RegisteredNumber, CarType, ManufacturerID, ManufacturerName, RenterID, RenterName, RenterAddress, Date, Time)

   a) non first NF

   b) $1^{st}$ NF

   c) $2^{nd}$ NF

   d) $3^{rd}$ NF

   e) BCNF

   f) $4^{th}$ NF

5. Onto which tables can you decompose the following table?

   Rent_a_car (RegisteredNumber, CarType, ManufacturerID, ManufacturerName, RenterID, RenterName, RenterAddress, Date, Time)

   a) Rent_a_car (RegisteredNumber, Date, Time)

   b) Cars (RegisteredNumber, CarType)

- c) Rent_a_car (RegisteredNumber, Date, Time, RenterID)
- d) Cars (RegisteredNumber, CarType, ManufacturerID)
- e) Renters (RenterID, RenterName, RenterAddress)
- f) Manufacturers (ManufacturerID, ManufacturerName)

6. After decomposition what will be the primary keys in the tables?

   | | |
   |---|---|
   | Cars | ManufacturerID |
   | Rent_a_car | RenterID |
   | Manufacturers | RegisteredNumber |
   | Renters | RegisteredNumber, Date |

7. After decomposition which columns do also appear as foreign key in the tables?

   a) RegisteredNumber
   b) RenterID
   c) ManufacturerID
   d) Date

**Normalization by Cookbook Method - Example**

Let us see the next table – we printed the primary key in bold!

T (**A**, B, C, D, **E**, F, G, H, I, J, K, L, M, N)

In the table T the following functional dependencies exist:

| | | | | |
|---|---|---|---|---|
| (2) | A, E → B, C, D, F, G, H, I, J, K, L, M, N | (11) | | |
| (1) | A → B, C, D | (3) | | 2$^{nd}$ NF |
| (1) | E → F, G, H, I, J, K, L, M | (8) | | 2$^{nd}$ NF |
| (1) | J → K, L, M | (3) | | 3$^{rd}$ NF |
| (2) | D, E → N | (1) | | BCNF |

We counted the attributes on left and right hand side and wrote the number beside the sides. In the right end column we indicated which normal forms are violated by functional dependencies.

The first and second dependencies have identical attributes on the right hand side. We keep these attributes in second dependency because the LHS of it is lower we delete these attributes from the first dependency. We proceed similarly with the attributes appearing in the first and the third functional dependencies. The result is:

| | | |
|---|---|---|
| (2) | A, E → ~~B, C, D, F, G, H, I, J, K, L, M~~, N | (11) |
| (1) | A → B, C, D | (3) |
| (1) | E → F, G, H, I, J, K, L, M | (8) |
| (1) | J → K, L, M | (3) |
| (2) | D, E → N | (1) |

The third and fourth dependencies have identical attributes on the right hand side. The LHS is one in both dependencies thus the attributes remain where the RHS is lower in which is in the fourth dependency, we delete them from third dependency.

The numbers of the attributes on left hand sides of first and fifth dependencies are the same, the N attribute appear on right hand side of both dependencies. Since RHS value of first dependency is higher we delete N attribute from it. The result is:

(2)     A, E → ~~B, C, D, F, G, H, I, J, K, L, M,~~ ~~N~~     (11)

(1)     A → B, C, D     (3)

(1)     E → F, G, H, I, J, ~~K, L, M~~     (8)

(1)     J → K, L, M     (3)

(2)     D, E → N     (1)

We transcribe the dependencies into tables and designate the primary and foreign keys. We printed the primary keys in bold and foreign keys in italics.

T (**A**, **E**)

T2 (**A**, B, C, D)

T3 (**E**, F, G, H, I, *J*)

T4 (**J**, K, L, M)

T5 (**D**, *E*, N)

## Authors

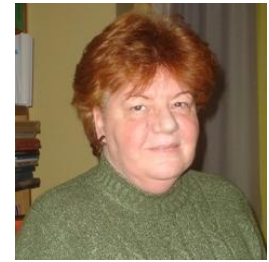

Márta Czenky is an assistant professor in the Department of Informatics of Faculty of Mechanical Engineering of Szent István University (Gödöllő, Hungary). She is teaching informatics, programming, database management, database programming in BSC and MSC qualifications at more faculties. Her research area is teaching of database management, her publications deal mainly with this topic. More books of her were published in Hungary, among others: *Data modeling, application of SQL and Access, Let's learn informatics together, Programming in Access, Exercise book of data modeling tasks, Exercise book of Excel tasks*.